\documentclass[aps,pra,twocolumn,showpacs,superscriptaddress,eqsecnum]{revtex4}
\usepackage{epsfig}
\usepackage{graphicx}
\usepackage{bm}
\begin{document}
\date{\today}
\title{Programmable unknown quantum-state discriminators with multiple copies of program and data: A Jordan basis approach}

\author{Bing He} \email{bhe98@earthlink.net}
\affiliation{Department of Physics and Astronomy, Hunter College of the City University of New York, 
 695 Park Avenue, New York, NY 10021}
 
\author{J\'{a}nos A. Bergou}
\affiliation{Department of Physics and Astronomy, Hunter College of the City University of New York, 
 695 Park Avenue, New York, NY 10021}


\begin{abstract}
The discrimination of any pair of unknown quantum states is performed by devices processing three parts of inputs: 
copies of the pair of unknown states we want to discriminate are respectively stored in two program systems and copies of data, 
which is guaranteed to be one of the unknown states, in a third system. We study the efficiency of such programmable 
devices with the inputs prepared with $n$ and $m$ copies of unknown qubits used as programs and data, respectively. 
By finding a symmetry in the average inputs, we apply the Jordan basis method to derive their optimal unambiguous 
discrimination and the minimum-error discrimination schemes. The dependence of the optimal solutions 
on the {\it a prior} probabilities of the mean input states is also demonstrated. 
\end{abstract}

\date{\today}
\maketitle

\section{Introduction} \label{section1}

The discrimination of quantum states is a nontrivial problem since a quantum state cannot be cloned perfectly if it 
is unknown to us \cite{Wootters82}. There are strategies of reaching the optimal results in the discrimination measurements: 
the minimum error probability discrimination \cite{Helstrom76,holevo}, where each measurement outcome select
one of the possible states and the error probability is minimized, and the optimal unambiguous discrimination
for the linearly independent states \cite{ivanovic,dieks,peres,jaeger,chefles98}, where unambiguity is paid
by the possibility of getting inconclusive results from the measurement. Recently an analogue of the optimal 
unambiguous discrimination is proposed for the linearly dependent state sets as the maximum confidence measurement \cite{c-e-b}.
In all these approaches, the input states to be discriminated occur or are prepared with some {\it a prior} probabilities 
in reality.

If the states we want to discriminate are known, we can use the given information to find the optimal measurements, which
can be either von Neumann measure or the generalized quantum measure in the form of positive operator value measure (POVM), 
to discriminate the elements in the set of states. If we know nothing about these 
states, however, the only information available will be the permutation symmetry with respect 
to a given number of copies of the unknown states provided to us. In the original work of two completely unknown states 
discrimination without ambiguity \cite{bergou-05}, the authors used a sort of programmable quantum devices, which have been 
studied both theoretically and experimentally in the recent years \cite {Nielsen97, Vidal00, Hillery, dusek, fiurasek-1,  
fiurasek-2,s-c-f-d, d-p}, to relate the program part of the inputs in a simple way to the unknown qubits $|\psi_{1}\rangle$ and 
$|\psi_{2}\rangle$ that one is trying to identify. The total input states
measured by the device are thus prepared with the pairs of unknown qubits as $|\psi_{1}\rangle_{A}|\psi_{1}\rangle_{B}
|\psi_{2}\rangle_{C}$ and $|\psi_{1}\rangle_{A}|\psi_{2}\rangle_{B}|\psi_{2}\rangle_{C}$, and the optimal POVM for the unambiguous discrimination of the inputs is designed with the permutation symmetry 
of the program registers $A,C$ and the data register $B$. If the \emph{a priori} 
probabilities of the inputs are equal, the maximum average success probability of discriminating them can be as large as
$1/6$.

The most general problem of this type is
when we have $n_A$ copies of the state in the program system $A$, $n_C$ 
copies of the state in the other program system $C$, and $n_B$ copies of the 
state in the data system $B$. Then the task is to discriminate 
two input states
\begin{eqnarray}
|\Psi_{1}^{in}\rangle & = & |\psi_{1}\rangle_{A}^{\otimes 
n_A}|\psi_{1}\rangle_{B}^{\otimes n_B}
|\psi_{2}\rangle_{C}^{\otimes n_C} \ , \nonumber \\
|\Psi_{2}^{in}\rangle & = & |\psi_{1}\rangle_{A}^{\otimes 
n_A}|\psi_{2}\rangle_{B}^{\otimes n_B}
|\psi_{2}\rangle_{C} ^{\otimes n_C}\ 
\end{eqnarray}
with the minimum error or with the least inconclusive probabilities, if we apply the minimum-error or the unambiguous strategies, respectively.
We should optimally distinguish between the above inputs with respect to these cost functions, keeping in mind that one has no knowledge of $|\psi_{1}\rangle$ and 
$|\psi_{2}\rangle$. The optimal schemes, in which the multiple copies of program or 
data used in the inputs, were solved for the input states prepared with equal \emph{a priori} probabilities but 
with an arbitrary number of copies in the program registers ($n_A=n_C=n$, $n_B=1$) \cite{Hayashi05_estimation,hayashi2}, for the 
inputs with multiple copies of data ($n_A=n_C=1$, $n_B=n$)\cite{j-v-e-h-h}, and for the unambiguous state discriminator working in 
the whole range of the \emph{a priori} probabilities of the inputs with $n_A=n_C=n$ and $n_B=1$\cite{bhe06-1}. 
The programmable discriminator to unambiguously discriminate a pair of unknown input states 
prepared with single program and data copies ($n_A=n_B=n_C=1$) is also generalized in the minimax approach \cite {dariano} to that 
of $N$ states \cite {zy}.

In this paper we study the unknown qubits discrimination with the inputs prepared with $n$ program copies and $m$ data
copies ($n_A=n_C=n$, $n_B=m$). Since the efficiency in discriminating the averages of the inputs in Eq. (1.1) indicates the 
upper bounds we aim to approach in the discrimination of a pair of unknown states, we will study the problem of discriminating 
the averaged input states as in \cite {Hayashi05_estimation, hayashi2, j-v-e-h-h} by both unambiguous and minimum-error 
discrimination strategies and generalize part of the results in these papers.
We will also show the dependence of the optimal solutions on the {\it a prior} 
probabilities of the averaged input states. To apply the Jordan basis method \cite {j-e-m06} to the optimal discriminations of the mean input states 
$<|\Psi_{1}^{in}\rangle\langle\Psi_{1}^{in}|>$ and $<|\Psi_{2}^{in}\rangle\langle\Psi_{2}^{in}|>$, we give a 
systematic study of the structure of these mean input states through finding an 
inherent symmetry which exists only under the condition $n_A=n_C$. In the limit of very large numbers of both program and  
data copies, we demonstrate that the definite discrimination with a unit success probability for the average input states will be 
approached for all \emph{a priori} probabilities. 

This paper is organized as follows: in Section II we demonstrate the equivalence between the discrimination of the average 
input states and that of uniformly distributed mixed states. The structure of the average input mixed states is discussed in
Section III and Appendix A and B, and the derivations of the inner products of the Jordan basis, which are used to represent the mean input states,
and their multiplicities are given in Section IV. Our main results about the optimal unambiguous discrimination and the minimum-error discrimination
of the average input states are given in Section V and VI, respectively. Finally, we give some conclusive discussion in the
last section.

\section{Mean Inputs as Uniformly Distributed Mixed States}\label{section2}

In this section we demonstrate that the discrimination of the average input states is equivalent to that of two uniformly distributed
mixed states. First we use the binomial expansion of $n$ copy tensor product 
of qubits,
\begin{widetext}
\begin{eqnarray}
(\cos(\theta/2 )|0\rangle+\sin (\theta/2)
e^{i\phi}|1\rangle)^{\otimes n} 
= \sum\limits_{k=0}^{n}
\cos^{n-k}(\theta/2)\sin^{k}(\theta/2)e^{ik\phi}\sqrt{C^k_n}|e_k\rangle\ ,
\end{eqnarray}
to introduce in the orthonormal basis $\{|e_i\rangle\}$:
\begin{eqnarray}
|e_0\rangle &=& \frac{1}{\sqrt{C^0_n}}|0,0,\ldots,0\rangle \nonumber\\
|e_1\rangle &=& \frac{1}{\sqrt{C^1_n}}(|1,0,\ldots,0\rangle+|0,1,\ldots,0\rangle+\cdots+|0,0,\ldots,1\rangle)\nonumber\\
&&\cdots  \nonumber\\
|e_k\rangle&=& \frac{1}{\sqrt{C^k_n}}(\underbrace{\underbrace{|1,1,\ldots,0,0\rangle}\limits_{k's~1~in~n~digits}+
|0,1,1,\ldots,,0\rangle +\cdots+|0,0,\ldots,1,1\rangle}\limits_
{summation~of~C^k_n~terms})\nonumber\\
&&\cdots\nonumber\\
|e_n\rangle &=&\frac{1}{\sqrt{C^n_n}}|1,1,\ldots, 1\rangle\ ,
\end{eqnarray}
\end{widetext}
where $C^k_n$ is the number of ways to choose $k$ objects from a group of $n$ objects without regard to order.
The index $k$ of $|e_k\rangle$ means how many digits $1$ in this basis vector out of the total digits,
and these basis vectors are orthonormal, $\langle e_i|e_j\rangle=\delta_{i,j}$.

As in the related works, 
we assume that the unknown qubits $|\psi_{1}\rangle$ and $|\psi_{2}\rangle$ uniformly and independently distribute over the Bloch 
spheres, and the averages of the inputs
with $n_A=n_C=n$ and $n_B=m$ are given as
\begin{eqnarray}
\rho_1 &=& \frac{1}{(4\pi)^2}\int d\psi_1\int d\psi_2 |\Psi_{1}^{in}\rangle\langle \Psi_{1}^{in}|\nonumber\\
&=&\frac{1}{(n+1)(n+m+1)}\sum\limits_{i=1}^{(n+1)(n+m+1)}|v_i\rangle\langle v_i|\nonumber\\
\rho_2&=&\frac{1}{(4\pi)^2}\int d\psi_1\int d\psi_2 |\Psi_{2}^{in}\rangle\langle \Psi_{2}^{in}|\nonumber\\
&=&\frac{1}{(n+1)(n+m+1)}\sum\limits_{i=1}^{(n+1)(n+m+1)}|v'_i\rangle\langle v'_i|,~~~~
\end{eqnarray}
where $|v_i\rangle \equiv |e_j\rangle_{A,B}\otimes |e_k\rangle_{C}$ ($0\leq j\leq n+m$, $0\leq k\leq n$) and 
$|v'_i\rangle \equiv |e_j\rangle_{A}\otimes |e_k\rangle_{B,C}$ ($0\leq j\leq n$, $0\leq k\leq n+m$),
and we have also used the integral,
\begin{eqnarray}
2\int_{0}^{\frac{\pi}{2}}sin^{2m-1}x~cos^{2n-1}x~dx=\frac{\Gamma(m)\Gamma(n)}{\Gamma(m+n)}.
\end{eqnarray}

In taking these averages we actually realize the one-to-one maps 
from the {\it unknown} qubit ensembles $\{|\psi_{1}\rangle\}$, $\{|\psi_{2}\rangle\}$ (the wave bracket means a set)
to two {\it known} mixed states:
\begin{eqnarray}
\{|\psi_1\rangle\}&\longmapsto & \rho_1 \nonumber\\
\{|\psi_2\rangle\}&\longmapsto & \rho_2.
\end{eqnarray}
Generally these mixed states are produced from the unknown qubit ensembles with some 
different {\it a prior} probabilities $\eta$ and $1-\eta$, respectively.
In the following we will apply the Jordan basis method  \cite {j-e-m06} to derive the optimal schemes 
for the unambiguous and the minimum-error discrimination of the mixed states $\rho_1$ and $\rho_2$.

\section{Structure of Mean Input States}\label{section3}

\subsection{Closed Chains of Basis Vectors}

We here study the structure of the mean input states $\rho_1$ and $\rho_2$. Let $H_1$ be the Hilbert space of $\rho_1$, which is spanned 
by $\{|v_i\rangle\}$, and $H_2$ the Hilbert 
space of $\rho_2$, which is spanned by $\{|v'_i\rangle\}$. We have $dim H_1\cap H_2=2n+m+1$, and the 
dimension of the total Hilbert space $H$ is therefore
\begin{eqnarray}
dim H&=&dim H_1\cup H_2=dim H_1+dim H_2-dim H_1\cap H_2 \nonumber\\
&=&2n^2+2nm+2n+m+1.
\end{eqnarray}

For a particular $|v_i\rangle$ in $H_1$, we need to find out with which
elements of $\{|v'_j\rangle\}$ spanning $H_2$ it has non-zero overlaps , i.e. to find out all $|v'_j\rangle$'s 
such that $\langle v_i|v'_j\rangle\neq 0$. To do it, we use the fact,
\begin{eqnarray}
C^k_{n+m}=C^{i}_{n} C^{j}_{m}+C^{i+1}_{n} C^{j-1}_{m}+C^{i+2}_{n} C^{j-2}_m+\cdots+C^{i+l}_{n} C^{j-l}_m,~~
\end{eqnarray}
where $i+j=k$,
to split $|e_j\rangle_{A,B}$ in $|v_i\rangle$ and
$|e_k\rangle_{B,C}$ in $|v'_i\rangle$ into the summation of tensor products by parts as follows:
\begin{widetext}
\begin{eqnarray}
|e_k\rangle_{A,B}=\sqrt{\frac{C^i_n C^j_m}{C_{n+m}^k}}|e_i\rangle_A |e_j\rangle_B +\sqrt{\frac{C^{i+1}_n C^{j-1}_m}{C_{n+m}^k}}
|e_{i+1}\rangle_A |e_{j-1}\rangle_B+\cdots+\sqrt{\frac{C^{i+l}_n C^{j-l}_m}{C_{n+m}^k}}|e_{i+l}\rangle_A |e_{j-l}\rangle_B
\end{eqnarray}
\end{widetext}
where $i+j=k$, and the number $l$ is determined by how many ways two non-negative integers can be summed up to be $k$.

Next, we use the $[n_1,n_2,n_3]$ symbol defined as follows to classify the mutually overlapped subspaces of $H_1$ and $H_2$. 
With the help of the above formula, we find that any basis vector $|v_p\rangle$ in $H_1$ or $|v'_p\rangle$ in $H_2$ has a 
unique representation in terms of this symbol:
\begin{widetext}

\begin{eqnarray}
|v_p\rangle&=&\sqrt{\frac{C^i_n C^j_m}{C_{n+m}^{i+j}}}|e_i\rangle_A |e_j\rangle_B |e_k\rangle_{C}+
\sqrt{\frac{C^{i+1}_n C^{j-1}_m}{C_{n+m}^{i+j}}}
|e_{i+1}\rangle_A |e_{j-1}\rangle_B |e_k\rangle_{C}+\cdots+\sqrt{\frac{C^{i+l}_n C^{j-l}_m}{C_{n+m}^{i+j}}}|e_{i+l}\rangle_A 
|e_{j-l}\rangle_B|e_k\rangle_{C}\nonumber\\
&\equiv & [i,j,k]+[i+1,j-1,k]+\cdots +[i+l,j-l,k],
\end{eqnarray}
\begin{eqnarray}
|v'_p\rangle&=&\sqrt{\frac{C^i_n C^j_m}{C_{n+m}^{i+j}}}|e_k\rangle_{A}|e_i\rangle_B |e_j\rangle_C +
\sqrt{\frac{C^{i+1}_n C^{j-1}_m}{C_{n+m}^{i+j}}}
|e_k\rangle_{A}|e_{i+1}\rangle_B |e_{j-1}\rangle_C +\cdots+\sqrt{\frac{C^{i+l}_n C^{j-l}_m}{C_{n+m}^{i+j}}}|e_k\rangle_{A}|e_{i+l}\rangle_B |e_{j-l}\rangle_C \nonumber\\
&\equiv & [k,i,j]+[k,i+1,j-1]+\cdots +[k,i+l,j-l] ,
\end{eqnarray}

\end{widetext}
where the coefficients of in each terms of the basis vectors have been absorbed into the square brackets.
Obviously we see from this expression that any couple of $|v_p\rangle$ and $|v'_p\rangle$ satisfy
$\langle v'_p|v_p\rangle\neq 0$ only if they have a common $[n_1,n_2,n_3]$ term.
 
By the $[n_1,n_2,n_3]$ representation of $|v_p\rangle$ or $|v'_p\rangle$, each term of basis vector has
the same number, $N\equiv n_1+n_2+n_3$, which goes from $0$ to $2n+m$. We'll show that
each distinct $N$ corresponds to a pair of closed basis vector chains, one of which is in $H_1$ and the other of which
in $H_2$, and their elements may have non-zero overlaps. 

Let's first look at $N=0$ and $N=2n+m$, the two simplest cases. $[0,0,0]=|0,0,\cdots,0\rangle$
and $[n,m,n]=|1,1,\cdots,1\rangle$ are shared by $H_1$ and $H_2$, and their overlaps are just $\langle v'|v\rangle=1$.
They are two initial pairs of closed basis vector chains with only one element.

When $N=1$, let's pick out one basis, $[0,1,0]+[1,0,0]$, in $H_1$ (by $[n_1,n_2,n_3]$ representation all the terms of a basis in 
$H_1$ have the same last digit $n_3$),
and then one basis, $[0,1,0]+[0,0,1]$, in $H_2$ (the first digit $n_1$ are the same 
in $[n_1,n_2,n_3]$ representation), which is overlapped with it.
The only other basis vector in $H_1$, with which $[0,1,0]+[0,0,1]$ of $H_2$ also has overlap, is $[0,0,1]$, while $[0,1,0]+[1,0,0]$ in
$H_1$ is also overlapped with $[1,0,0]$ in $H_2$. Thus we exhausted all basis vectors in $H_1$ and $H_2$ with $N=1$ and obtain such
a pair of closed chains as follows:
\begin{eqnarray}
|v_1\rangle &=&[0,1,0]+[1,0,0],\nonumber\\
|v_2\rangle &=& [0,0,1];
\end{eqnarray}
\begin{eqnarray}
|v'_1\rangle &=&[0,1,0]+[0,0,1],\nonumber\\
|v'_2\rangle &=& [1,0,0].
\end{eqnarray}
Both of them have $2$ elements and, by retriving the coefficients absorbed in the square brackets, we can easily
find all their overlaps:
\begin{eqnarray}
\langle v'_1 |v_1\rangle &=& \frac{C^1_m}{C^1_{n+m}},\nonumber\\
\langle v'_1 |v_2\rangle &=& \langle v_1 |v'_2\rangle =\sqrt{\frac{C^1_n}{C^1_{n+m}}}.
\end{eqnarray}

For $N=2n+m-1$, we find in the same way the following two closed chains:
\begin{eqnarray}
|v_1\rangle &=&[n,m-1,n]+[n-1,m,n],\nonumber\\
|v_2\rangle &=& [n,m,n-1];
\end{eqnarray}
\begin{eqnarray}
|v'_1\rangle &=&[n,m-1,n]+[n,m,n-1],\nonumber\\
|v'_2\rangle &=& [n-1,m,n];
\end{eqnarray}
and, with the combinatorics identities, we find that they have the same overlaps as in $N=1$ case, so we call them 
the conjugate chains of $N=1$.

As $N$ increases to any $ i\leq n $, the closed chains of basis vectors can be found inductively in the above way:
\begin{eqnarray}
|v_1\rangle &=&[0,i,0]+[1,i-1,0]+[2,i-2,0]+\cdots +[i,0,0],\nonumber\\
|v_2\rangle &=& [0,i-1,1]+[1,i-2,1]+\cdots +[i-1,0,1],\nonumber\\
|v_3\rangle &=& [0,i-2,2]+\cdots +[i-2,0,2]\nonumber\\
&&\cdots \nonumber\\
|v_{i+1}\rangle &=& [0,0,i];
\end{eqnarray}
\begin{eqnarray}
|v'_1\rangle &=&[0,i,0]+[0,i-1,1]+[0,i-2,2]+\cdots +[0,0,i],\nonumber\\
|v'_2\rangle &=& [1,i-1,0]+[1,i-2,1]+\cdots +[1,0,1-1],\nonumber\\
|v'_3\rangle &=& [2,i-2,0]+\cdots +[2,0,i-2]\nonumber\\
&&\cdots \nonumber\\
|v'_{i+1}\rangle &=& [i,0,0].
\end{eqnarray}
For every $|v_j\rangle$ ($j\leq i+1$) in a chain of $H_1$, the last index $n_3=j-1$ increases as $j$ from $0$
to $i$, while
for every $|v'_j\rangle$ ($j\leq i+1$) in a chain of $H_2$, the first index $n_1=j-1$ increases as $j$ from $0$
to $i$. The number 
of different terms in a $|v_j\rangle$ (resp. $|v'_j\rangle$) is how many ways the non-negative integers $n_1$ and
$n_2$ (resp. $n_2$ and $n_3$) can be added
up to $i-j+1$, so it goes down from $i+1$ to $1$ as $j$ increases. 

If $N=2n+m-i$, there are two chains conjugate to those of $N=i$: the last index $n_3$ of $|v_j\rangle$ decreases
from $n$ to $n-i$ as $j$ goes up, while the first index $n_1$ of $|v'_j\rangle$ also decreases
from $n$ to $n-i$ as $j$ goes up. The number 
of different terms in a $|v_j\rangle$ (resp. $|v'_j\rangle$) is how many ways the non-negative integers $n_1$ and
$n_2$ (resp. $n_2$ and $n_3$) can be added
up to $n+m-i+j-1$.

In the pairs of closed chains with mutual overlaps, whenever $N$ increases its values by $1$, the number of 
the elements in the closed chains will increase by $1$, if $0\leq N\leq n$. On the other hand, the number of 
the elements in closed chains also increases by $1$ as $N$ decreases by $1$, if $n+m\leq N\leq 2n+m$. However, the last index
$n_3$ of the $|v_i\rangle$ in $H_1$ and the first index
$n_1$ of the $|v'_i\rangle$ in $H_2$ cannot increase beyond $n$ or decrease beyond $0$, 
so the number of elements in a closed chain cannot be larger than $n+1$.

The size of a closed chain thus increases in two directions from both $N=0$ and $N=2n+m$, 
and it will be fixed with $n+1$ elements as $N$ increases to the medium value $n$ or decreases to the medium value $n+m$. 
The number of such chains with the maximum $n+1$ elements in $H_1$ and $H_2$ is $m+1$. Therefore, the total number of the elements
in all closed chains in $H_1$ or $H_2$ equals
\begin{eqnarray}
dim H_1&=&dimH_2=2\sum_{i=1}^{n}i+(m+1)(n+1)\nonumber\\
&=&(n+m+1)(n+1).
\end{eqnarray}

After we have obtained all pairs of closed chains with mutual overlaps this way, we will find that
the $\{|v_i\rangle\}$ in a closed chain of $H_1$ and its counterpart $\{|v'_i\rangle\}$ in $H_2$ can be permutated 
such that they have symmetric overlap, $\langle v'_i|v_j\rangle =\langle v_i |v'_j\rangle$, for any couple of $i$ and $j$
(we can also have $i=j$ because the inner products are real). We call it a {\it mirror symmetry} \cite{symmetry}, and give a 
procedure of how to obtain pairs of closed chains with this symmetry in Appendix A. This symmetry leads to very useful invariants
for finding the Jordan basis inner products for each pair of closed basis vector chains.

\subsection{Mean Input States Represented by the Jordan Basis}

As it is proved for any pair of subspaces \cite {jordan},
there exist the Jordan basis $\{|\phi_i\rangle\}$ in $H_1$ and $\{|\phi'_i\rangle\}$ in $H_2$ , with which the mean input
states $\rho_1$ and $\rho_2$ are expressed as 
\begin{eqnarray}
\rho_1&=&\frac{1}{(n+1)(n+m+1)}\sum\limits_{i=1}^{(n+1)(n+m+1)}|\phi_i\rangle\langle \phi_i|\nonumber\\
\rho_2&=&\frac{1}{(n+1)(n+m+1)}\sum\limits_{i=1}^{(n+1)(n+m+1)}|\phi'_i\rangle\langle \phi'_i|,~~~~~~
\end{eqnarray}
where $\langle\phi_i|\phi'_j\rangle=0$ for each couple of $i\neq j$. If we have obtained the mean states with the above 
representation, we will be able to use the strategy in Ref. \cite{j-e-m06} to get the optimal scheme for the unambiguous
discrimination of them.

To get the mean inputs represented by the Jordan basis, we need to look for the orthogonal transformations,
\begin{eqnarray}
\left(\begin{array}{c}|\phi_1\rangle\\
 |\phi_2\rangle\\ \vdots\\|\phi_k\rangle
\end{array}\right) = \left(\begin{array}{cccc}
a_{11}& a_{12}&\cdots & a_{1k} \\
a_{21}&a_{22}&\cdots & a_{2k} \\
\vdots&\vdots &\ddots & \vdots \\
a_{k1}&a_{k2}&\cdots & a_{kk}\\
\end{array}\right)
\left(\begin{array}{c}|v_1\rangle\\
 |v_2\rangle\\ \vdots\\|v_k\rangle
\end{array}\right)
\end{eqnarray}
in $H_1$, and the corresponding orthogonal transformations,
\begin{eqnarray}
\left(\begin{array}{c}|\phi'_1\rangle\\
 |\phi'_2\rangle\\ \vdots\\|\phi'_k\rangle
\end{array}\right) = \left(\begin{array}{cccc}
a'_{11}& a'_{12}&\cdots & a'_{1k} \\
a'_{21}&a'_{22}&\cdots & a'_{2k} \\
\vdots&\vdots &\ddots & \vdots \\
a'_{k1}&a'_{k2}&\cdots & a'_{kk}\\
\end{array}\right)
\left(\begin{array}{c}|v'_1\rangle\\
 |v'_2\rangle\\ \vdots\\|v'_k\rangle
\end{array}\right)
\end{eqnarray}
in $H_2$, for all pairs of closed basis vector chains (the index $k$ means the size of a closed chain).
Though it is very difficult to obtain all these orthogonal transformations analytically, in the following we will use the
symmetric property of the basis vector chains to derive all the Jordan basis inner products, $\langle\phi'_i|\phi_i\rangle$, 
for each closed basis vector chain, which are enough for us to find the optimal POVMs and their success probabilties for the 
unambiguous and the minimum-error discriminations of our average input states.

By the inverse orthogonal transformations, $|v_i\rangle=\sum_{k}a_{ki}|\phi_k\rangle$ and $|v'_i\rangle=\sum_{k}a'_{ki}|\phi'_k\rangle$, 
of Eqs. (3.15)-(3.16), we deduce from the {\it mirror symmetry}, $\langle v'_i|v_j\rangle =\langle v_i |v'_j\rangle$, the following 
equations,
\begin{eqnarray}
\sum\limits_k a_{ik}a'_{kj}\langle\phi_k|\phi'_k\rangle
=\sum\limits_k a'_{ik}a_{kj}\langle\phi'_k|\phi_k\rangle,
\end{eqnarray}
for each pair of $i=j$ or $i\neq j$. We have 
the sufficient condition, $a_{ij}=a'_{ij}$ for all $i$'s and $j$'s, to guarantee the validity of all these equations and,
therefore, two orthogonal transformations in Eqs. (3.15)-(3.16) can be reduced to a single one in realizing the transformation
to the Jordan basis.

Substituting $|\phi_i\rangle=\sum_{k}a_{ik}|v_k\rangle$ and 
$|\phi'_i\rangle=\sum_{k}a_{ik}|v'_k\rangle$ into $\sum_i \langle\phi_i|\phi'_i\rangle$ and using the orthogonal transformation 
property $\sum_i a_{k'i}a_{ik}=\delta_{kk'}$, we will find the following
invariants,
\begin{eqnarray}
S_N&=&\sum_{i=0}^N\langle\phi_i|\phi'_i\rangle =\sum _{i=0}^{N} \left((\sum_{k'} \langle v_{k'}|a_{k'i})
(\sum_k a_{ik}|v'_k\rangle)\right) \nonumber\\
&=&\sum_{i=0}^N\langle v_i|v'_i\rangle,
\end{eqnarray}
if $0\leq N\leq n-1$. For the conjugate chains with $n+m+1\leq N'\leq 2n+m$ we have $S_{N'}=S_{2n+m-N}$,
and the $m+1$ pairs of closed chains with the maximum $n+1$ elements all have the same invariant.
These invariants are very useful in finding all the Jordan basis inner products, and
they are determined by how many elements in a closed basis vector chain. We give a
simple method of how to calculate these invariants with the $[n_1,n_2,n_3]$ symbol in Appendix B.

\section{Derivation of Jordan Basis Inner Products and Their Multiplicities}\label{section4}

For $N=0$ and $N=2n+m$ the Jordan basis are just the original ones, and the inner product is obviously
$\langle\phi_1|\phi'_1\rangle=1$, i.e. they belong to the intersection of two subspaces.

In $N=1$ (similarly in $N=2n+m-1$) case, we apply the rotation
$T_{1,2}$ on the closed chains given by Eqs (3.6)-(3.7),
\begin{eqnarray}
\left(\begin{array}{c}|\phi_1\rangle\\
 |\phi_2\rangle\\
\end{array}\right)&=& T_{1,2}\left(\begin{array}{c}|v_1\rangle\\
 |v_2\rangle\\
\end{array}\right)
=\left(\begin{array}{cc}\cos\theta & \sin\theta\\
 -\sin\theta & \cos\theta\\
\end{array}\right)
\left(\begin{array}{c}|v_1\rangle\\
 |v_2\rangle\\
\end{array}\right), \nonumber\\
\left(\begin{array}{c}|\phi'_1\rangle\\
 |\phi'_2\rangle\\
\end{array}\right)&=& T_{1,2}\left(\begin{array}{c}|v'_1\rangle\\
 |v'_2\rangle\\
\end{array}\right)
=\left(\begin{array}{cc}\cos\theta & \sin\theta\\
 -\sin\theta & \cos\theta\\
\end{array}\right)
\left(\begin{array}{c}|v'_1\rangle\\
 |v'_2\rangle\\
\end{array}\right),~~~~~~
\end{eqnarray}
to obtain $\{|\phi_1\rangle,|\phi_2\rangle \}$ and $\{|\phi'_1\rangle,|\phi'_2\rangle \}$.  
The conditions
\begin{eqnarray}
\langle \phi_1|\phi'_2\rangle=\langle \phi'_1|\phi_2\rangle=0,
\end{eqnarray}
imply
\begin{eqnarray}
-\langle v_1|v'_1\rangle \tan\theta+\langle v_1|v'_2\rangle (1-\tan^2\theta)=0,
\end{eqnarray}
a quadratic equation that can be solved easily.
Substituting $\tan\theta$ and $\langle v_1|v'_1\rangle$, $\langle v_1|v'_2\rangle$ into the inner products of Jordan basis, 
we obtain
\begin{eqnarray}
\langle \phi_1|\phi'_1\rangle &=& 1,\nonumber\\
\langle \phi_2|\phi'_2\rangle &=& -\frac{n}{n+m}.
\end{eqnarray}
We can verify that their summation equals to the invariant $S_1$, which is the square of the coefficient of $[0,1,0]$ term, $\frac{C^1_m}{C^1_{n+m}}$.

The $N=2$ Jordan basis inner products are obtained by first applying two successive rotations $T=T_{1,3}T_{1,2}$ in $3$ dimensional 
subspace and then imposing the vanishing inner products for the pairs of rotated vectors with different indices to 
seperate one couple of basis vectors from others; after that we use $T_{2,3}$ 
to seperate the rest pairs in the same way. The rotation $T_{i,j}$ 
in the subspace of a chain is defined as an $n\times n$ ($n$ is the number
of chain elements) identity matrix with the elements in the
positions $\{i,i\}$, 
$\{i,j\}$, $\{j,i\}$ and $\{j,j\}$ replaced by the corresponding elements of an $O(2)$ matrix. Following the procedure,
we obtain the inner products with their absolute values arranged in a descending order:
\begin{eqnarray}
\langle \phi_1|\phi'_1\rangle &=& 1,\nonumber\\
\langle \phi_2|\phi'_2\rangle &=& -\frac{n}{n+m},\nonumber\\
\langle \phi_3|\phi'_3\rangle &=& \frac{n(n-1)}{(n+m)(n+m-1)}.
\end{eqnarray}
The summation of these Jordan basis inner products is equal to the invariant $S_2$, which can be independently obtained
by summing up the square of the $[0,2,0]$ coefficient, $\frac{C_m^2}{C_{n+m}^2}$, and the square of $[1,0,1]$ coefficient,
$\frac{C^1_n}{C_{n+m}^1}$, if $m\geq 2$. If $m< 2$ instead, we use only $[1,0,1]$ term to find the same
result for $m=1$.

We see from the above results that the first two inner products of $N=2$ chains are those of $N=1$ chains, and the first inner product
is just that of $N=0$ chains, so this recurrence gives rise to the multiplicities of Jordan basis inner products in the whole Hilbert space. 

For any pair of closed chain with a fixed $N$ number, if we apply successive orthogonal transformations, 
$T_1=T_{1,N}T_{1,N-1}\cdots T_{1,3}T_{1,2}$, $T_2=T_{2,N}T_{2,N-1}\cdots T_{2,3}$, $\cdots$, and impose the vanishing inner 
products for the pairs of basis vectors with different indices, we will obtain the Jordan basis inner products as in the above 
equations. A useful obsevation in doing so is the recurrence of the Jordan basis inner products from closed chains
to closed chains, which is found by induction as we have done for the chains of $N=0,1,2$. The general recurrent pattern for these
Jordan basis inner products is as follows: the inner products in the chains of $N=k$ coincide with those of $N=k+1$ chains except
for the one extra inner product of $N=k+1$, if $0\leq k\leq n-1$; meanwhile, the inner products of $N=k$ chains are those of $N=k-1$
except for the one extra inner product of $N=k-1$, if $n+m+1\leq k \leq 2n+m$.

Together with the invariants $S_N$, we can tremendously simplify the calculation to get the Jordan basis inner products with this
recurrence. If we have obtained the Jordan basis inner products for the chains of $N=k$
($0\leq k\leq n-1$), the one more unsolved inner product for the chains of $N=k+1$ is just $S_{k+1}-S_k$, which can be independently 
obtained from the closed chains represented by the original basis $\{|v_i\rangle\}$ and $\{|v'_i\rangle\}$. Starting from $N=0$, we 
can find all the Jordan basis inner products for the chains of $0\leq N\leq n-1$ this way. For example, the extra Jordan basis inner
product for the $N=3$ chains is obtained by $S_3$ and Eq. (4.5) as follows:
\begin{eqnarray}
&\langle \phi_4|\phi'_4\rangle& = S_3-S_2\nonumber\\
&=&\frac{C_m^3}{C_{n+m}^3}+\frac{C_n^1C_m^1}{C_{n+m}^2}-\frac{C_m^2}{C_{n+m}^2}-
\frac{C^1_n}{C_{n+m}^1}\nonumber\\
&=&-\frac{n(n-1)(n-2)}{(n+m)(n+m-1)(n+m-2)},
\end{eqnarray}
where $S_3$ is from the contribution of $[0,3,0]$ and $[1,1,1]$ terms, if we suppose $m\geq 3$ (otherwise we can check
out the above result too with the relevant terms for $m=1$ and $m=2$, respectively). For the chains of $n+m+1\leq N \leq 2n+m$, we 
can use the similar procedure to get all their Jordan basis inner products too. Therefore, the $m+1$ pairs of closed chains
with the constant invariants $S_N$ for $n\leq N\leq n+m$ have the following complete set of $n+1$ Jordan basis products:
\begin{eqnarray}
\langle \phi_1|\phi'_1\rangle &=& 1,\nonumber\\
\langle \phi_2|\phi'_2\rangle &=& -\frac{n}{n+m},\nonumber\\
\langle \phi_3|\phi'_3\rangle &=& \frac{n(n-1)}{(n+m)(n+m-1)},\nonumber\\
&&\cdots\nonumber\\
\langle \phi_{n+1}|\phi'_{n+1}\rangle &=& \pm \frac{n(n-1)\cdots 1}{(n+m)(n+m-1)\cdots (m+1)}.~~~~~~~~
\end{eqnarray}
The sign for the last inner product is determined by whether $n$ is even or odd. Thus we have exhausted
all the closed basis chains and obtained all their Jordan basis products.

As the conclusion for this section, we list all inner products of Jordan basis and their multiplicities in the following table:
\begin{center}
\doublerulesep 2.5pt
\begin{tabular}{|c|c|}\hline\hline
inner product & multiplicity\\\hline
1& 2n+m+1\\\hline
$-\frac{n}{n+m}$& $2n+m-1$ \\\hline
$\frac{n(n-1)}{(n+m)(n+m-1)}$ &$2n+m-3$\\\hline
$-\frac{n(n-1)(n-2)}{(n+m)(n+m-1)(n+m-2)}$ &$2n+m-5$\\\hline
$\vdots$ & $\vdots$\\\hline
$\pm \frac{n(n-1)\cdots 1}{(n+m)(n+m-1)\cdots (m+1)}$&$m+1$\\\hline
\end{tabular}
\end{center}
The inner products, $\langle \phi|\phi'\rangle=1$, correspond to the intersection 
of $H_1$ and $H_2$, and we also see that each closed chain in $H_1$ has a one-dimensional joint space with the 
corresponding chain in $H_2$.

\section{Optimal Unambiguous Discrimination of Average Input States}\label{section5}

In Ref. \cite{j-e-m06}, the optimal scheme for the unambiguous discrimination of two mixed states represented by the Jordan basis, 
\begin{eqnarray}
\rho_1&=&\sum\limits_i\alpha_i|\phi_i\rangle\langle \phi_i|\nonumber\\
\rho_2&=&\sum\limits_i\beta_i|\phi'_i\rangle\langle \phi'_i|,
\end{eqnarray}
which occur with the {\it a prior} probability of $\eta$ and $1-\eta$, respectively, 
is derived through optimizing the POVM in the form of $\Pi_{1}=\sum_{i=1}^{k}
\Pi_{1,i}$, $\Pi_{2}=\sum_{i=1}^{k}\Pi_{2,i}$ and $\Pi_0=I-\Pi_1-\Pi_2$ with $I$ being the identity operator,
where 
\begin{equation}
\Pi_{1,i}=\frac{1-q_{i}}{1-|\langle\phi'_i|\phi_i\rangle|^2}|z_{i}\rangle\langle z_{i}|,
\end{equation}
and
\begin{equation}
\Pi_{2,i}=\frac{1-q'_{i}}{1-|\langle\phi'_i|\phi_i\rangle|^2}|y_{i}\rangle\langle y_{i}|.
\end{equation}
The orthonormal basis $\{|z_{i}\rangle\}$ and $\{|y_{i}\rangle\}$ in the above equations satisfy
\begin{eqnarray}
|\langle\phi_i|z_{i}\rangle|^2&=&1-|\langle\phi'_i|\phi_i\rangle|^2,~~~~~~~\langle\phi'_j|z_{i}\rangle=0,\nonumber\\
|\langle\phi'_i|y_{i}\rangle|^2&=&1-|\langle\phi'_i|\phi_i\rangle|^2,~~~~~~~\langle\phi_j|y_{i}\rangle=0, 
\end{eqnarray}
for all $i$'s and $j$'s, and $q_{i}$, $q'_{i}$ the failure probabilities in the unambiguous discrimination of couples of
$|\phi_i\rangle$ and $|\phi'_i\rangle$ for all $i$'s. Then the total failure probability for the discrimination of two mixed states,
\begin{eqnarray}
Q_L  &= &\sum_{i=1}^{k} Q_{i}\nonumber\\
&=&  \sum_{i=1}^{k}[\eta \alpha_{i}q_{i}(\eta) 
+(1-\eta) \beta_{i} q'_i(\eta)] \ ,
\end{eqnarray}
is optimized with $Q_{i}$ taking the following solutions:
\begin{equation}
Q_{i}^{opt} = \left\{\begin{array}{ll} \eta \alpha_{i}+(1-\eta)
\beta_{i}|\langle\phi'_i|\phi_i\rangle|^2 & \mbox{if $\eta \leq c_{i}$}\\
 2\sqrt{\eta(1-\eta)\alpha_{i}\beta_{i}}|\langle\phi'_i|\phi_i\rangle| & \mbox{if $c_{i} \leq \eta \leq d_{i}$}\\
 \eta\alpha_{i}|\langle\phi'_i|\phi_i\rangle|^2+(1-\eta)\beta_{i} & \mbox{if $\eta \geq d_{i}$}\end{array} \right. \ ,
 \label{Qiopt}
\end{equation}
where $c_i$, $d_i$ are the boundaries of the range given as
\begin{equation}
I_i=[c_{i},d_{i}]=\left[ \frac{\beta_{i} |\langle\phi'_i|\phi_i\rangle|^2}{\alpha_{i}+
\beta_{i} |\langle\phi'_i|\phi_i\rangle|^2}, \frac{\beta_{i}}{\beta_{i} +
\alpha_{i} |\langle\phi'_i|\phi_i\rangle|^2} \right] .
\end{equation}
In the range of $I_{0}=\bigcap_{i}I_{i}$, where the total POVM is valid, the optimal POVM taking all absolute minimums on the 
center line of Eq. (5.6) achieves the success probability,
\begin{eqnarray}
P(\eta)=1-Q_L^{opt}=1-2\sqrt{\eta(1-\eta)}\sum_i\sqrt{\alpha_i\beta_i}|\langle\phi_i|\phi'_i\rangle|,~~~~
\end{eqnarray}
if the spaces of two mixed sates to be discriminated join at the origin. 

In our problem, however, there is a non-trivial intersection space of the Hilbert spaces $H_1$ and $H_2$, so this intersection 
contributes to failure probability and should be added to the total failure probability $Q_L$ too. With all the Jordan basis inner 
products and their multiplicities derived in the last section, we can readily obtain the optimal scheme
for the unambiguous discrimination of our average input states prepared with any number of program and data.

Taking the $\alpha_i$ and $\beta_i$ in Eq. (3.14) in the general situation with the input states produced with $n$ copies of 
program and $m$ copies of data, we find the validity range, 
\begin{eqnarray}
I_0=\left[\frac{n^2}{(n+m)^2+n^2},\frac{(n+m)^2}{(n+m)^2+n^2}\right],
\end{eqnarray}
and the optimal success probabilty,
\begin{widetext}
\begin{eqnarray}
P(\eta)&=& 1-Q_L^{opt}\nonumber\\
&=&1-\frac{2n+m+1}{(n+m+1)(n+1)}
-2\sqrt{\eta(1-\eta)}\frac{1}{(n+m+1)(n+1)}\{\frac{n(2n+m-1)}{n+m}+ \frac{n(n-1)(2n+m-3)}{(n+m)(n+m-1)}\nonumber\\
&&+\frac{n(n-1)(n-2)(2n+m-5)}{(n+m)(n+m-1)(n+m-2)}
 + \cdots+  \frac{n(n-1)\cdots 1}{(n+m)(n+m-1)\cdots (m+2)} \},
\end{eqnarray}
of the total POVM.
We have
\begin{eqnarray}
P(\eta) < 1-\frac{2n+m+1}{(n+m+1)(n+1)} -\frac{2m+2}{(n+m+1)(n+1)}
&\times& K(n,m)\sqrt{\eta(1-\eta)},
\end{eqnarray}
and
\begin{eqnarray}
P(\eta) > 1-\frac{2n+m+1}{(n+m+1)(n+1)}
-\frac{4n+2m-2}{(n+m+1)(n+1)}\times K(n,m)\sqrt{\eta(1-\eta)},
\end{eqnarray}
where 
\begin{eqnarray}
K(n,m)=\frac{n}{n+m}+  \frac{n(n-1)}{(n+m)(n+m-1)}+\frac{n(n-1)(n-2)}{(n+m)(n+m-1)(n+m-2)}
+\cdots +\frac{n(n-1)\cdots 1}{(n+m)(n+m-1)\cdots (m+1)}~~~~~~
\end{eqnarray}
\end{widetext}
is a typical convergent series.
From these inequalities, we see that $P(\eta)$ will tend to a certain $1$ if both $n$ and $m$ tend to infinity.

We here give some examples when the optimal success probabilities $P(\eta)$ can be reduced to the closed forms:

For one copy program and $m$ copies data in the input states, we have
\begin{eqnarray}
P(\eta)=1-\frac{m+3}{2(m+2)}-\frac{1}{m+2}\times \sqrt{\eta(1-\eta)};
\end{eqnarray}
For $2$ copies of program and $m$ copies of data used in the input states, we get
\begin{eqnarray}
P(\eta)=1&-&\frac{m+5}{3(m+3)}\nonumber\\
&-&\frac{4(m+4)}{3(m+2)(m+3)}\times \sqrt{\eta(1-\eta)};~~~~
\end{eqnarray}
For $n$ copies program and $1$ copy data in the input states, the optimal success probability is found as follows:
\begin{eqnarray}
P(\eta)&=&1-\frac{2}{n+2}-
\frac{2}{(n+1)(n+2)}\times2\sqrt{\eta(1-\eta)}\nonumber\\
&&\times \frac{1}{n+1}(n^2+(n-1)^2+\cdots+2^2+1^2)\nonumber\\
&=&1-\frac{2}{n+2}-\frac{2n(2n+1)}{3(n+1)(n+2)}\times \sqrt{\eta(1-\eta)}.~~~~~~~~~~
\end{eqnarray}
If $\eta=0.5$ (equal {\it a prior} probabilities), it reduces to
\begin{eqnarray}
P=\frac{n}{3(n+1)}
\end{eqnarray}
given in \cite{hayashi2} and tends to $1/3$, the optimal IDP average \cite {ivanovic, dieks, peres} in discriminating a pair of 
known qubits, as $n$ goes to infinity.

\section{Minimum-error Discrimination of Average Input States}\label{section6}

The general problem of discriminating two mixed states $\rho_1$ and $\rho_2$, which occur with different {\it a prior}
probabilities $\eta_1$ and $\eta_2$, respectively, with the minimum of the error probability,
\begin{eqnarray}
P_E=\eta_1 Tr(\rho_1\Pi_2)+\eta_2 Tr(\rho_2\Pi_1),
\end{eqnarray}
is solved by classifing the enginevalue spectrum $\{\lambda_k\}$ of the operator \cite{Helstrom76},
\begin{eqnarray}
\Lambda=\eta_2\rho_2-\eta_1\rho_1=\sum\limits_{k=1}^{dim H}\lambda_k |\omega_k\rangle\langle\omega_k|.
\end{eqnarray}
The corresponding optimal measurement is achieved by two projectors,
\begin{eqnarray}
\Pi_1&=&\sum\limits_{k=1}^{k_0-1}|\omega_k\rangle\langle\omega_k|\nonumber\\
\Pi_2&=&\sum\limits_{k=k_0}^{dim H}|\omega_k\rangle\langle\omega_k|,
\end{eqnarray}
where $\Pi_1$ is the projector onto the space spanned by $k_0-1$ eigenvectors of $\Lambda$ with negative eigenvalues and $\Pi_2$ 
the projector onto the space spanned by the remaining eigenvectors with non-negative eigenvalues, and they
satisfy $\Pi_1+\Pi_2=I$, where $I$ is the identity operator.
The minimum error probability is therefore given as
\begin{eqnarray}
P_E=\frac{1}{2}(1-Tr|\Lambda|)=\frac{1}{2}(1-Tr|\eta_2\rho_2-\eta_1\rho_1|),
\end{eqnarray}
with $|\Lambda|=\sqrt{\Lambda^{\dagger}\Lambda}$.

With the Jordan basis representation of our mean input states, the corresponding $\Lambda$ operator is given as 
\begin{eqnarray}
\Lambda&=&\frac{1}{(n+1)(n+m+1)}\sum_{i=1}^{(n+1)(n+m+1)}\Lambda_i,
\end{eqnarray}
where $\Lambda_i=\eta_2|\phi'_i\rangle\langle \phi'_i|-\eta_1|\phi_i\rangle\langle \phi_i|$. We derive its eigenvalue spectrum
in the following.

The subspace spanned by each pair of $\{|\phi_i\rangle,|\phi'_i\rangle\}$, where $1\leq i\leq (n+1)(n+m+1)$, 
is orthogonal to the rest part of Hilbert space $H$. In it we introduce in the following new bases:
\begin{eqnarray}
|\omega_i\rangle=|\phi_i\rangle=|\phi'_i\rangle,
\end{eqnarray}
if $\langle \phi_i|\phi'_i\rangle=1$;
and 
\begin{eqnarray}
|\omega_i\rangle&=&\frac{1}{\sqrt{2(1+\langle \phi_i|\phi'_i\rangle)}}(|\phi_i\rangle+|\phi'_i\rangle),\nonumber\\
|\omega'_i\rangle&=&\frac{1}{\sqrt{2(1-\langle \phi_i|\phi'_i\rangle)}}(|\phi_i\rangle-|\phi'_i\rangle),
\end{eqnarray}
if $\langle \phi_i|\phi'_i\rangle\neq 1$. All these new bases are orthonormal ($\langle \omega_i|\omega_j\rangle=\delta_{i,j}$, $\langle \omega'_i|\omega'_j\rangle=\delta_{i,j}$ and
$\langle \omega_i|\omega'_j\rangle=0$).

With this new basis, the $\Lambda_i$ in the second case is given as
\begin{eqnarray}
\Lambda_i
&=&\left(\begin{array}{cc}\langle\omega_i|\Lambda_i|\omega_i\rangle & \langle\omega_i|\Lambda_i|\omega'_i\rangle \\
 \langle\omega'_i|\Lambda_i|\omega_i\rangle  & \langle\omega'_i|\Lambda_i|\omega'_i\rangle \\
\end{array}\right)\nonumber\\
&=&\left(\begin{array}{cc}\frac{1}{2}(\eta_2-\eta_1)(1+\kappa_i) &-\frac{1}{2}\sqrt{1-\kappa_i^2}\\
-\frac{1}{2}\sqrt{1-\kappa_i^2}&\frac{1}{2}(\eta_2-\eta_1)(1-\kappa_i)
\end{array}\right),~~~~~~
\end{eqnarray}
where $\kappa_i\equiv \langle \phi_i|\phi'_i\rangle$.
We thus obtain the following eigenvalues of the $\Lambda_i$:
\begin{eqnarray}
\lambda^{(i)}_1&=&\frac{1}{2}(c+\sqrt{1-(1-c^2)\kappa_i^2})\nonumber\\
\lambda^{(i)}_2&=&\frac{1}{2}(c-\sqrt{1-(1-c^2)\kappa_i^2}),
\end{eqnarray}
where $c\equiv \eta_2-\eta_1$. The eigenvalue spectrum of $\Lambda$ is therefore obtained as follows:
\begin{widetext}
\begin{eqnarray}
\Lambda &=&\frac{1}{(n+1)(n+m+1)}\sum\limits_{i=1}^{2n+m+1}c|\omega_i\rangle\langle\omega_i|
+\frac{1}{2(n+1)(n+m+1)}\sum\limits_{i=2n+m+2}^{(n+m+1)(n+1)}\left(c+\sqrt{1-(1-c^2)\kappa_i^2}\right)
|\lambda_i\rangle\langle\lambda_i|\nonumber\\
&+&\frac{1}{2(n+1)(n+m+1)}\sum\limits_{i=2n+m+2}^{(n+m+1)(n+1)}\left(c-\sqrt{1-(1-c^2)\kappa_i^2}\right)
|\lambda'_i\rangle\langle\lambda'_i|,~~~~~~~~~
\end{eqnarray}
\end{widetext}
where $|\lambda_i\rangle$ and $|\lambda'_i\rangle$ are the eigen-vectors corresponding to the eigenvalues
$\lambda^{(i)}_1$ and $\lambda^{(i)}_2$, respectively.

Together with the table of Jordan basis inner products $\langle \phi_i|\phi'_i\rangle$ in Section IV, this eigenvalue 
spectrum of $\Lambda$ allows us to obtain the minimum error probability for arbitrary $n$ and $m$ in the input states.
Here we give two examples when $c=0$ (equal preparation probabilities $\eta_1=\eta_2$) and compare their results with 
those of optimal unambiguous discrimination we obtained previously.

First, we take $1$ copy of program and $n$ copies of data. In this case there is a multiplicity of $n+1$ for 
$\langle \phi_i|\phi'_i\rangle=-1/(n+1)$ and none of other Jordan basis inner product is not equal
to $1$. Plugging these results into (6.4), we obtain  
\begin{eqnarray}
P_E=\frac{1}{2}\left(1-\frac{1}{2}\sqrt{\frac{n}{n+2}}~\right).
\end{eqnarray}
As $n$ goes to infinity, $P_E$ tends to $1/4$. In the unambiguous discrimination for this case, the least failure probability
$Q_L$ has the limit of $1/2$ as $n$ tends to infinity. So we have the relation, $P_E=0.5 Q_L$, if we have $1$ copy of program and 
infinite copies of data in our inputs.

Another example is to have $n$ copies of program and $1$ copy data.  After substituting
the Jordan basis inner products and the corresponding multiplicites into the Helstrom formula Eq. (6.4), we obtain the minimum 
error probability
\begin{eqnarray}
P_E= \frac{1}{2}\left( 1-\frac{2}{n+2}\sum\limits_{i=1}^{n}\sqrt{1-(\frac{i}{n+1})^2}~\frac{i}{n+1} \right).~~~~
\end{eqnarray}
If $n\rightarrow \infty$, the limit of $P_E$ is
\begin{eqnarray}
P_E=\frac{1}{2}\left (1-2\int_{0}^{1} \sqrt{1-x^2}~x~ dx \right)=\frac{1}{6},
\end{eqnarray}
which is consistent with the result obtained by other method in \cite{Hayashi05_estimation} together with those of all other
$n$. Compared with the corresponding least failure probability, $Q_L=2/3$, in the unambiguous discrimination, 
we have $P_E=\frac{1}{4}Q_L$ if we are dealing with the inputs carrying infinite copies of program and $1$ copy of data.

The general relation \cite{b-h-m}, $P_E\leq \frac{1}{2}Q_L$,
for the error probability $P_E$ and the failure probability $Q_L$ of 
the unambiguous state discrimination is well satisfied for these extreme situations. Moreover, we see from the results
a prominent difference of the minimum-error discrimination from the unambiguous discrimination: more copies of program copies will give 
higher success probability, while in the unambiguous discrimination more data copies gives higher success probability.

\section{Concluding Remarks}\label{section7}
So far we have studied the problem of discriminating pairs of unknow qubits with the devices processing the inputs prepared with
$n$ and $m$ copies of them used for the programs and the data, respectively.
The upper bound efficiency of the devices are achieved by optimally discriminating two uniformly distributed mixed states 
if we assume the pair of qubits we want to discriminate uniformly and independently distribute over their Bloch spheres. 
We show that there exists a symmetry in these mixed states when number of state copies in two program registers are equal. 
If we represent these mixed states with the Jordan basis, we can conveniently obtain the Jordan basis inner 
products and their corresponding multiplicities with this symmetry. Then we have the optimal schemes of the unambiguous 
and the minimum-error discrimination of these mean inputs occuring with arbitrary {\it a prior} probabilities. 

We should also compare the upper bound efficiency of the unknown state discriminators we discussed with that in discriminating a pair of known 
pure states $|\psi_1\rangle$ and $|\psi_2\rangle$. Given $m$ copies of such known states, the 
upper bound success probability \cite {ivanovic, dieks, peres}, $1-|\langle\psi_1|\psi_2\rangle|^m$, for their unambiguous 
discrimination, which is lower than that of the minimum-error discrimination, 
promises a definite identification if $m$ tends to infinity. Corresponding to our unknown input states with $m$ copies of data 
in Eq. (1.1), however, we see that the 
definite discrimination can be approached only if both $n$ and $m$ are very large numbers. In the design of any programmable 
discriminator for unknown states, we will need to use large program registers to improve its efficiency.

For the situation of arbitrary  $n_A$, $n_B$ and $n_C$, the {\it mirror symmetry} we find in our mean 
input states is gone, so it is impossible to obtain the analytical solutions to the optimal discrimination problems 
as what are discussed in this paper. In this even more general situation we need to deal with the mixed states of different 
dimensions. With the structure of the mean input states we clarify in this paper, however, a numerical algorithm can 
be worked out to obtain their Jordan basis inner products, since there exist the Jordan basis for a pair of arbitrary subspaces 
\cite {jordan}. It is also possible to generalize the optimal solutions to those of unknown {\it qudits} discrimination. 
These problems will be studied in the future work.

\noindent 

\appendix

\section{Procedure to Obtain Symmetric Closed Basis Vector Chains}

Here we give a procedure of obtaining the closed basis vector chains with the {\it mirror symmetry},
$\langle v_i|v'_j\rangle=\langle v'_i|v_j\rangle$, for any couple of indices $i$ and $j$. 

Step 1: Generating the closed basis vector chains of any fixed $N\equiv n_1+n_2+n_3$. 

In $H_1$, since each term of a basis vector in this representation has the same digit $n_3$, we start from $n_3=0$
and obtain the corresponding basis vector (that is to find out all the combinations of non-zero intergers that 
added up to be $N$). 
Then we increase $n_3$ by $1$ each time till $n_3=n$ and obtain in this way all basis vectors in this closed chain. 
Similarly we obtain the corresponding chain in $H_2$, each element of which has the same first digit $n_1$ for all terms.

Step 2: Permutation of the basis vectors, which is given by $[n_1,n_2,n_3]$ representation, in the obtained closed basis chains.

Remember that $[i,j,k]\equiv \sqrt{\frac{C^i_n C^j_m}{C_{n+m}^{i+j}}}|e_i\rangle_A |e_j\rangle_B |e_k\rangle_{C}$ in $H_1$
and $[k,j,i]\equiv \sqrt{\frac{C^i_n C^j_m}{C_{n+m}^{i+j}}}|e_k\rangle_A |e_j\rangle_B |e_i\rangle_{C}$ in $H_2$.
Pick out in a closed chain all couples of basis vectors carrying $[l_1,l_2,l_3]$ and $[l_3,l_2,l_1]$ terms (the terms with
$n_1$ and $n_3$ interchanged), which are 
guranteed to exist by the fact that $N$ is constant for any of a closed chain, and let the basis vector with $[l_1,l_2,l_3]$
be $|v_i\rangle$, and that with $[l_3,l_2,l_1]$ be $|v_{i+k}\rangle$, where $i$, $k$ are the integers within the range of the size
for a closed chain, in $H_1$. Correspondingly in $H_2$, we set the basis vector carrying $[l_3,l_2,l_1]$ instead be $|v'_i\rangle$,
and that carrying $[l_1,l_2,l_3]$ be $|v'_{i+k}\rangle$. 

Then we proceed from other terms in the basis vectors to obtain two whole
chains with the {\it mirror symmetry}.
If the number of program copies $n_A$ and $n_C$ are unequal, we cannot guarantee the existence
of the couples in the form $\{[l_1,l_2,l_3],[l_3,l_2,l_1]\}$, and the {\it mirror symmetry} of the basis vectors doesn't exist.

With a not so large $N$, we here give an example of how to perform the procedure. 
For the case of $4$ copies of program and $1$ copy of data in the input
states, suppose that we need to obtain the chains of $N=4$ with the {\it mirror symmetry}. 

There are $5$ elements in the chains as we studied previously.
Starting from $n_3=0$ and $n_1=0$, we see that the couple $\{[l_1,l_2,l_3],[l_3,l_2,l_1]\}$ in the basis vectors
is $\{[4,0,0],[0,0,4]\}$. Adding the rest terms in these basis vectors, we set $|v_1\rangle=[4,0,0]+[3,1,0]$, $|v_2\rangle=[0,0,4]$ in $H_1$, and
$|v'_1\rangle=[0,0,4]+[0,1,3]$, $|v'_2\rangle=[4,0,0]$ in $H_2$. These two couples of basis satisfy
$\langle v_1|v'_2\rangle=\langle v'_2|v_1\rangle=\sqrt{\frac{C^4_4C^0_1}{C^4_5}}\sqrt{\frac{C^0_4C^0_1}{C^0_5}}$.

The basis in $H_2$ carrying $[3,1,0]$ is $[3,0,1]+[3,1,0]$, and we set it as $|v'_3\rangle$. In $H_1$ the basis carrying
$[0,1,3]$ is $[1,0,3]+[0,1,3]$, which is set as $|v_3\rangle$. We have
$\langle v_1|v'_3\rangle=\langle v'_1|v_3\rangle=\sqrt{\frac{C^3_4C^1_1}{C^4_5}}\sqrt{\frac{C^0_4C^1_1}{C^1_5}}$. Then we
set $|v_4\rangle=[3,0,1]+[2,1,1]$ and $|v'_4\rangle=[1,0,3]+[1,1,2]$, and there is 
$\langle v_3|v'_4\rangle=\langle v'_3|v_4\rangle=\sqrt{\frac{C^1_4C^0_1}{C^1_5}}\sqrt{\frac{C^3_4C^0_1}{C^3_5}}$.

Finally we consider the terms with $[2,1,1]$ and $[1,1,2]$; in $H_1$ we set $|v_5\rangle=[2,0,2]+[1,1,2]$ and in $H_2$ we
set $|v'_5\rangle=[2,0,2]+[2,1,1]$, and 
$\langle v_4|v'_5\rangle=\langle v'_4|v_5\rangle=\sqrt{\frac{C^2_4C^1_1}{C^3_5}}\sqrt{\frac{C^1_4C^1_1}{C^2_5}}$. 

Thus we obtain the following two closed chains having the {\it mirror symmetry}. 
\begin{eqnarray}
|v_1\rangle&=&[4,0,0]+[3,1,0],\nonumber\\
|v_2\rangle&=&[0,0,4],\nonumber\\
|v_3\rangle&=&[1,0,3]+[0,1,3],\nonumber\\
|v_4\rangle&=&[3,0,1]+[2,1,1],\nonumber\\
|v_5\rangle&=&[2,0,2]+[1,1,2];
\end{eqnarray}
\begin{eqnarray}
|v'_1\rangle&=&[0,0,4]+[0,1,3],\nonumber\\
|v'_2\rangle&=&[4,0,0],\nonumber\\
|v'_3\rangle&=&[3,0,1]+[3,1,0],\nonumber\\
|v'_4\rangle&=&[1,0,3]+[1,1,2],\nonumber\\
|v'_5\rangle&=&[2,0,2]+[2,1,1].
\end{eqnarray}

Even if $N$ is very large, we can search out all these couples with symmetric inner product and perform 
the above procedure by a computer program.

\vspace{0.2cm}

\section{Calculation of Invariant Summation of Basis Vector Inner Products}

With the $[n_1,n_2,n_3]$ symbol defined in section III, it is very easy to calculate the invariants, 
$S_N=\sum_i\langle v_i|v'_i\rangle=\sum_i\langle\phi_i|\phi'_i\rangle$, for the pairs of closed basis vector 
chains, given arbitrary $n$ program and $m$ data copies in the input states.
In a pair of closed chains with the {\it mirror symmtry}, only the terms in the form of  $[i,j,i]$ ($n_1=n_3$) contribute to this 
invariant sum, because $|v_i\rangle$'s and $|v'_i\rangle$'s carrying the terms in all other forms are orthogonal, so it is obtained 
by picking out the terms of $[i,j,i]$ satisfying $i\leq n$, $j\leq m$ and summing up the square of the coefficients absorbed in these square brackets. It can be done easily by a computer program for any pair of closed 
chains. 

Here we use the situation of $4$ copies of program and $3$ copies of data in the input states for an example. 
We calculate the invariant sums for $N=4,5,6,7$ chains with $5$ elements together as shown previously.

For $N=4$ the contribution to the sum come from the terms $[2,0,2]$ and $[1,2,1]$, so the invariant sum is
\begin{eqnarray}
\sum_i\langle\phi_i|\phi'_i\rangle=\sum_i\langle v_i|v'_i\rangle=\frac{C^2_4C^0_3}{C_7^2}+\frac{C_4^1C^2_3}
{C^3_7}=\frac{22}{35}.
\end{eqnarray}

For $N=5$ the contribution to the sum come from the terms $[2,1,2]$ and $[1,3,1]$, so the invariant sum is
\begin{eqnarray}
\sum_i\langle\phi_i|\phi'_i\rangle=\sum_i\langle v_i|v'_i\rangle=\frac{C^2_4C^1_3}{C_7^3}+
\frac{C_4^1C^3_3}
{C^4_7}=\frac{22}{35}.
\end{eqnarray}

For $N=6$ the contribution to the sum come from the terms $[2,2,2]$ and $[3,0,3]$, so the invariant sum is
\begin{eqnarray}
\sum_i\langle\phi_i|\phi'_i\rangle=\sum_i\langle v_i|v'_i\rangle=\frac{C^2_4C^2_3}{C_7^4}+
\frac{C_4^3C^0_3}
{C^3_7}=\frac{22}{35}.
\end{eqnarray}

For $N=7$ the contribution to the sum come from the terms $[3,1,3]$ and $[2,3,2]$, so the invariant sum is
\begin{eqnarray}
\sum_i\langle\phi_i|\phi'_i\rangle=\sum_i\langle v_i|v'_i\rangle=\frac{C^3_4C^1_3}{C_7^4}+
\frac{C_4^2C^3_3}
{C^5_7}=\frac{22}{35}.
\end{eqnarray}

All these pairs of chains with $5$ elements have the same invariant sum. By induction on $n$ and $m$, we see  that this invariant 
sum is determined by the size of the closed basis vector chains.

\vspace{0.2cm}

\bibliographystyle{unsrt}

\begin{thebibliography}{99}
\bibitem {symmetry} The prime sign over the shoulders of $v$'s is reflected by the perpendicular
line in the middle of the inner product like a mirror: $\langle v'_i|v_j\rangle =\langle v_i |v'_j\rangle$.
\end{thebibliography}

\end{document}